\documentstyle[epsfig,12pt]{article}
\def\circa#1{\,\raise.3ex\hbox{$#1$\kern-.75em\lower1ex\hbox{$\sim$}}\,}

\newcommand  \f  \varphi
\newcommand \bra {\langle}
\newcommand \ket {\rangle}
\newcommand{\be}{\begin{equation}}
\newcommand{\ee}{\end{equation}}
\newcommand{\ben}{\begin{displaymath}}
\newcommand{\een}{\end{displaymath}}
\newcommand{\ba}{\begin{eqnarray}}
\newcommand{\ea}{\end{eqnarray}}
\newcommand{\ban}{\begin{eqnarray*}}
\newcommand{\ean}{\end{eqnarray*}}

\newcommand{\g}{\gamma}
\newcommand{\kvet}{\bf k}

\begin{document}
\begin{titlepage}
\vspace{1cm}

\begin{center}
{\huge \bf Leading electroweak corrections}
\\
{\huge \bf at the TeV scale }
\\
\vspace{2cm}
{\Large{\bf P. Ciafaloni}}\\
\vspace{1cm}
{\it \large
Dipartimento di Fisica \& INFN, via Arnesano, 73100 Lecce \\
E-mail: Paolo.Ciafaloni@le.infn.it
}

\vspace{5cm}
{\large\bf Abstract}
\end{center}
\begin{quotation}
The planned next generation of linear colliders (NLCs) will be able to probe
the infrared  structure of Standard Model electroweak interactions, that
determines the behavior of electroweak radiative corrections
at TeV scale energies. I present results of a recent calculation at the
leading log level, and discuss my view on open issues and possible
future developments of this new and interesting subject.
\end{quotation}
\vspace{3cm}
\end{titlepage}

\def\baselinestretch{1.1}

\section{Introduction}
There has been recently an outburst of interest in the TeV scale 
behavior of Standard Model   electroweak corrections, triggered
by the observation that such behavior is dominated by the infrared 
(IR) structure of the theory \cite{ciafacome1}. 
The main motivation is of course  the possibility of having, 
in a hopefully nearby future,  
linear colliders operating at such very high energies and with high
luminosities \cite{NLC}. 
As has been pointed out in \cite{ciafacome1,beccariaetal}, at the TeV scale 
leading log (LL) one loop electroweak corrections have a typical magnitude of 
10 \% relative to the Born level. 
The reason for this\footnote{at LEP
electroweak corrections have a much smaller typical magnitude of 
$\frac{\alpha(M_Z)}{4 \sin^2\theta_w \pi}\approx 2.7\,\times\,10^{-3}$}
is that when the c.m. energy $\sqrt{s}$ is much bigger
than the electroweak scale, the W and Z masses act as effective cutoffs for
infrared divergences. One loop corrections then
grow like a {\sl double} logarithm of  $\sqrt{s}$, i.e.
like $\log^2\frac{\sqrt{s}}{M}$ 
where $M (\approx M_W\approx M_Z)$ is the electroweak scale. This
in contrast with the corrections related to the ultraviolet behavior
of the theory, described by the usual RGE equations and
growing like a {\sl single} log, being
therefore subdominant at very high energies.
Moreover, besides leading logs also
subleading effects of infrared origin are,
generically speaking,
numerically relevant \cite{beccariaetal}.
Since big SM effects could mask possible effects of New Physics at NLCs, an
accurate calculation of electroweak SM corrections is necessary, which in turn
implies addressing higher order calculations.
In addition, as I will try to show in the following, the IR structure of a
broken theory like the electroweak sector of the Standard Model is 
interesting {\sl in itself}, both from a theoretical and a phenomenological
point of view.

Three different calculations  \cite{Kuhn,ciafacome2,fadin}
of LL electroweak corrections for processes relevant at NLCs
have recently been done. These calculations give different results; 
this seems
to indicate that maybe some issues are still to be understood. In any case,
electroweak interactions have a distinctive feature that differentiates them
from both QED and QCD: symmetry breaking. The importance of this feature 
and its
relationship with the infrared structure of electroweak interactions are still
to be clarified. 

The main result of \cite{ciafacome2}, 
and also its main motivation,
is that, at a c.m. energy close to the TeV scale,
soft QED effects\footnote{here by ``soft QED effects'' I mean 
all photon  contributions giving LL effects}
cannot be accounted for separately as has been done 
in the LEP-era \cite{Altarelli:1989hv}. What happens instead, is that there is
a separation of scales such that below the electroweak scale $M$
only QED LL effects are present, 
while above $M$   the contributions of all  gauge
bosons $\g,W,Z$ to leading log effects have to
be taken into account together\footnote {as has also been noticed in
\cite{fadin}, 
in this region considering separately only the $W,Z$ contributions would
violate gauge invariance}. While the precise meaning of this 
sentence will be specified in next section, I would like to notice that the
above implies that the approach to radiative electroweak
corrections at NLCs must be
substantially changed from the one which was customary
 at LEP, nicely described in \cite{Altarelli:1989hv}.

Of course, since as I have said three different calculations give different
results, the issues commented above are currently
under debate. Eventually, 
a full, exact  two loop calculation without any {\it apriori} assumption
could discriminate between the different possibilities. 
In this case, I believe that  one
should consider   a physical process as simple as possible; the one we 
examine in \cite{ciafacome2} is a good example.

\section{The Z' electroweak form factor}
We have computed electroweak LL corrections at TeV scale with the following
approach in mind:
\begin{itemize}
\item
{\sl Complete} electroweak virtual corrections are calculated,
taking into account also the photon contribution. The photon is given a mass
$\lambda$ to regularize IR divergences.
\item
soft photon emission is calculated, with photons having energy less than
the experimental resolution ${\Delta E}$
\end{itemize}
The effect of soft  photon emission is basically 
that of substituting the photon
mass $\lambda$, which is an IR regulator,
  with the experimental resolution $\Delta E$. 
Therefore in the final result we give to $\lambda$ the physical meaning of 
an energy-angle experimental resolution parameter.
We are thinking about experimental resolutions of the order of 
$\lambda\approx$ 10
GeV, much lower than the W and Z bosons mass so that a process 
with W- or Z- bremsstrahlung is experimentally resolved.
In other words, we are inclusive with respect to emitted soft
photons, but exclusive with respect to W, Z emission.

In the spirit of choosing the simplest possible case for studying LL
electroweak
interactions at the TeV scale, we consider in \cite{ciafacome2} the two
fermions decay rate of a Z' gauge boson unmixed with the usual Z boson and
belonging to a group that commutes with the SM group. Since the Z' is neutral
under the Standard Model
 SU(2)$\otimes$U(1) gauge group, the relevant LL electroweak
corrections act only upon the two fermion external legs. Moreover, in the
massless fermion limit we consider, chirality is conserved and one can
consider separately the cases of left and right final fermions. Even though
this is one of the simplest cases of phenomenological interest one can
imagine, the basic formalism we set up and the general considerations we make
about the IR structure of electroweak SM interactions (i.e. factorization,
exponentiation and so on) are relevant for a more general class of processes
of interest at NLC energies. 

In order to compute the leading radiative corrections in the infrared region
$\sqrt{s}\gg w\gg M$, where $w$ is the virtual boson energy and $M$ its mass,
we use the method of soft insertions formulae, which are
widely used in QED \cite{low} and are known to provide 
in QCD \cite{cornwall,BCM}
the leading IR singularities at double log level. This method consists
in factorizing the softest virtual momentum $k_1=(w_1,\kvet)$ by computing 
external line insertions only.
The left-over diagram is then evaluated   by setting
$k_1=0$ (or, equivalently, the diagram is evaluated on-shell). 
This procedure is iterated and the final
integration is performed in the region of phase space giving rise to 
 the LL electroweak corrections we are interested in, i.e. 
the strongly ordered region: $w_1\ll w_2\ll.....\ll w_n$ where 1
labels the outermost boson and $n$ the innermost one (see fig. 1).
\begin{figure}[htb]\setlength{\unitlength}{1cm}
\begin{picture}(12,6)
\put(3.5,5.5){(a)}\put(8.5,5.5){(b)}\put(13.7,5.5){(c)}
\put(3.2,4){$W,Z,\g$}
\put(3,0){$Z'$}
\put(2,0){\epsfig{file=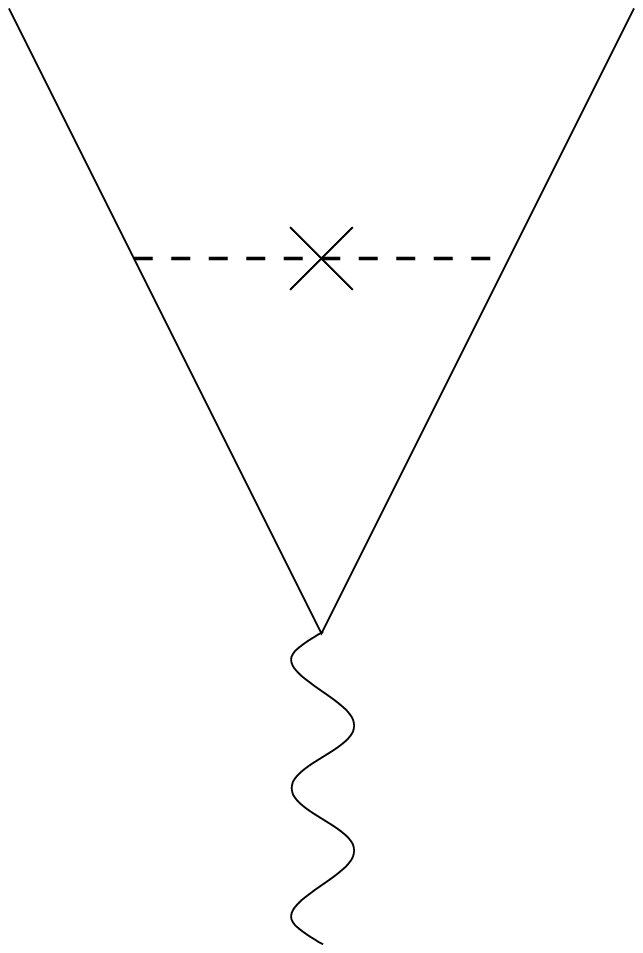,height=5cm}}
\put(6.5,4.5){$1$}
\put(7,3.5){$2$}
\put(8,2){$n$}
\put(8,0){$Z'$}
\put(7,0){\epsfig{file=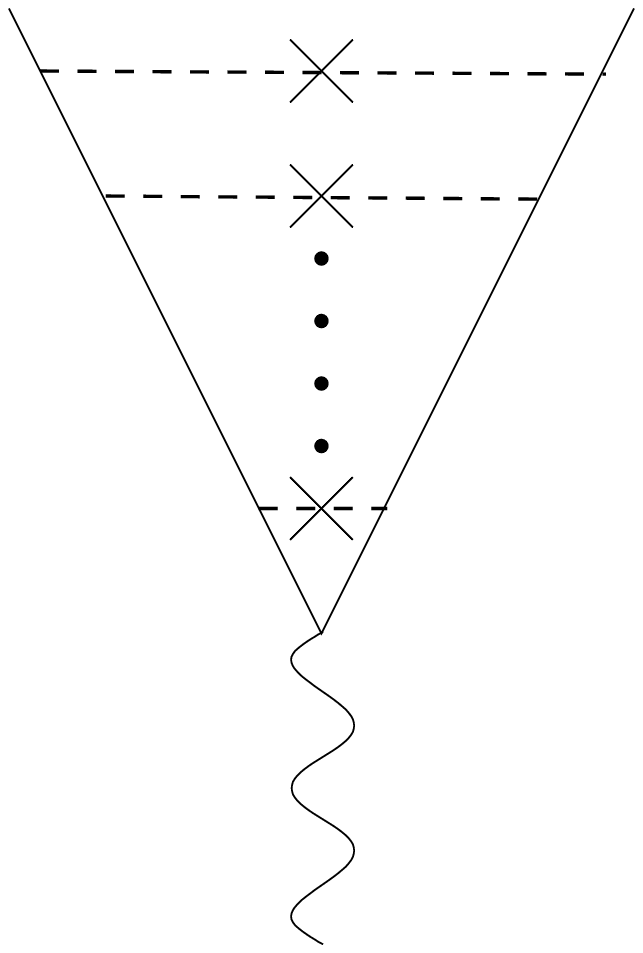,height=5cm}}
\put(12.4,4.5){QED}
\put(13.6,4.5){{\footnotesize {$\lambda<w<M$}}}
\put(12.4,3){EW}
\put(13.3,3){{\footnotesize {$M<w<\sqrt{s}$}}}
\put(12,0){\epsfig{file=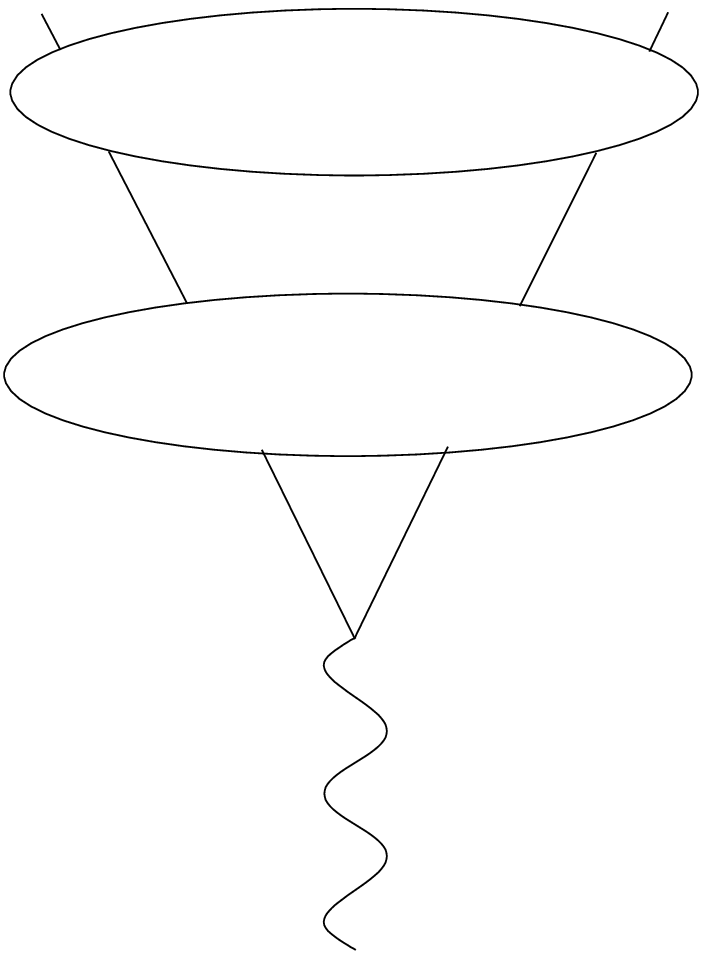,height=5cm}}
\end{picture}
\caption{(a)-(b)
:diagrams for soft boson insertion at 1 and n loops. Continuous lines
are fermion lines, and dashed lines are $W,Z,\g$ gauge bosons. Crosses
indicate that gauge bosons are close to mass-shell, and energies are such that
$w_1\ll w_2\ll.....\ll w_n$ (see text). (c): Pictorial representation of 
eq. (\ref{ff})}
\end{figure}

Referring to \cite{ciafacome2} for computational details and using here the
same notations, we find that the matrix element at the LL level  
is given by ($x\propto
\log w$ is used in place of $w$; $M_0$ is the Born level amplitude):
\be\label{ff}
{\bf M}^{LL}=\exp[-e^2q_f^2\frac{l^2}{2}]
\sum_{n=0}^\infty { M}_n=
\exp[-e^2q_f^2\frac{l^2}{2}]\bra f|P_x\exp[-\int_{0}^LdxH_{EW}(x)]|f\ket M_0
\ee\be\label{imp2}
H_{EW}(x)=
(g'^2Y\tilde{Y}+g^2\bar{T}\cdot\bar{\tilde{T}})x
+e^2Q\tilde{Q}l\qquad
l=\log\frac{M}{\lambda}\quad
L=\log\frac{\sqrt{s}}{M}\quad
x_i=\log\frac{w_i}{M}
\ee
Here $|f\ket$ is a fermion belonging to a given representation of
the SU(2)$\otimes$U(1) gauge group,  coupling with gauge bosons  
through  SU(2)$\otimes$U(1) 
generators $T^a$ normalized to  to Tr$\{T^aT^b\}=\frac{1}{2}\delta^{ab}$. 
The charge operator is defined as $Q=T^3+Y$\footnote{Operators 
are in capital letters, c-numbers in small
letters. Thus $Q$ is an operator with values $q_e=-1,q_\nu=0$}; 
tilded operators $\tilde{T^a}$ 
act on the
antifermion flavor indices\footnote{here and in the following,
by ``flavor'' I mean SU(2)$\otimes$U(1) quantum numbers; we only consider a
single generation of fermions} 
while untilded operators act on the fermion line. 
Furthermore, $P_x$ denotes the $x$-ordered product; 
at the LL level we can identify 
$M_Z\approx M_W\equiv M$ with the weak scale. 

Let us now discuss equation (\ref{ff})  in various regimes, by noting first
that  in this problem there are three relevant
 mass scales: the c.m. energy $\sqrt{s}$,
the electroweak scale $M$ and the parameter $\lambda$, which in the final
result has the meaning of an experimental resolution, as mentioned above.

First, when $\sqrt{s}\sim M$, only QED soft effects are present, which amounts
to saying that the ``LEP-approach'' is correct at LEP energies of course. 
However, for $\sqrt{s}\gg M\sim \lambda$ (or, equivalently, $l\ll L$)
 a completely different situation occurs and 
eqn. (\ref{ff}) then reduces to:
\be {\bf M}^{LL}=\exp[-(g'^2y_f^2+\frac{3}{4}g^2)\frac{1}{2}
\log^2\frac{\sqrt{s}}{M}] M_0
\ee
In other words, at very high energies what really factorizes and exponentiates
is the whole SU(2)$\otimes$U(1) group contribution, and \underline{not} 
the U(1)$_{em}$ component. This was to be expected, since the symmetry of
relevance to the problem is related to the energy scale. In particular, at
scales typical of electroweak symmetry breaking, one ``sees'' the pattern 
of breaking  SU(2)$\otimes$U(1)$\to$ U(1)$_{em}$, but when the energy gets
much bigger than this scale, the full gauge symmetry SU(2)$\otimes$U(1) is restored.

In general, and for arbitrary values of $\sqrt{s}$ above the electroweak
scale,  one
sees that QED effects are not completely factorized. In fact if this were the
case, then we would find a QED prefactor $\exp[-e^2q_f^2
\frac{1}{2}\log^2\frac{\sqrt{s}}{\lambda}]$
in place of  $\exp[-e^2q_f^2\frac{l^2}{2}]=
\exp[-e^2q_f^2\frac{1}{2}\log^2\frac{M}{\lambda}]$ in (\ref{ff}). 
In other words, only
photons with energies such that $\lambda<w_\g<M$ indeed factorize, 
as is depicted
pictorially in fig. 1c. For energies higher than $M$ the photon contribution
is taken into account in a nontrivial way in $H_{EW}$; 
this is precisely what we mean
by ``nonfactorizable soft QED effects''. As is shown in \cite{ciafacome2}, if
one insists in extrapolating to TeV scale energies
 the ``LEP approach'' in which soft QED corrections are
computed separately, then the error one makes with respect to the correct
result is at the LL level, i.e. an error precisely of the same order of the
effect  one is trying to calculate.

In second place, (\ref{ff},\ref{imp2}) 
are responsible for the fact that electroweak
LL effects do not exponentiate in a trivial way. In fact one finds the one and
2 loop results:
\be\label{results}
M_1=-(a_f\frac{L^2}{2}+b_flL)M_0\qquad
M_2=
\left\{\frac{1}{2}(a_f\frac{L^2}{2}+b_flL)^2
-\frac{1}{3}e^2g^2lL^3y_ft^3_f\right\}M_0
\ee
\be\label{eq:aandb}
a_f=g^2\frac{3}{4}+g'^2y_f^2\qquad\qquad
b_f=e^2q_f^2
\ee
and in the 2 loop result a ``spurious'' term $\propto e^2g^2 t^3$ is present. 
It is easy to show why there is no exponentiation in our approach. 
In fact, suppose that one sets the $Q\tilde{Q}$ term in (\ref{imp2}) to 0.
Then, what is left is the SU(2)$\otimes$U(1) Casimir, 
which is  a c-number. The
$x$ ordered exponential then simply produces  the regular exponential of
the abovementioned Casimir. However of course, the  $Q\tilde{Q}$ term can 
not be set to 0 in our approach: this is a noncommuting operator and
determines the terms that break exponentiation, that are proportional 
to $l$ as one can see from (\ref{results}). It is interesting to notice that
these effects come about because even for $w>M$, the Z and $\g$ bosons are
still distinguished by their masses, the latter acting as collinear
cutoffs. The ``exponentiation breaking'' term turns out to be proportional to 
$l=\log\frac{M}{\lambda}$.
Therefore without symmetry breaking, that causes mixing in the
neutral sector and gives rise to different mass scales for the 
gauge bosons, we
would have exponentiation. This is what happens for instance in QCD which
shares with the electroweak sector the property of being a nonabelian theory, 
but in which 
effects analogous to the ones we are studying do in fact exponentiate.

\section{Conclusions}

The bulk of radiative corrections at
energies much higher than the electroweak scale  is determined
by the infrared structure of the theory; this makes the subject a very
interesting (and experimentally testable!) one. 
The requirement of controlling Standard Model
contributions
at a level comparable to the experimental precision at NLCs 
implies addressing higher order calculations. 
The fact that the first calculations of this kind of  effects are
(at least partially) in disagreement, might indicate that there is 
a peculiar feature that
differentiates the  infrared structure of a broken theory like  the Standard
Model of 
electroweak interactions from the one of an unbroken theory like QCD 
for instance.
The main conclusion that emerges from our work \cite{ciafacome2}
is that the ``LEP-approach'' to electroweak corrections 
must be substantially changed, or improved, when considering electroweak
corrections at TeV scale energies.
From a theoretical point of view, in this problem
the interplay between symmetries of the theory and
energy scales at which the theory is tested is an important issue and in my
opinion, a not yet fully understood one.

\end{document}